\documentclass[preprint,review,12pt]{elsarticle}

\usepackage{amssymb}

\journal{Nuclear Physics B}

\begin{document}

\begin{frontmatter}



\title{
Critical behavior of the Higgs- and Goldstone-mass gaps
for the two-dimensional $S=1$ $XY$ model
}


\author{Yoshihiro Nishiyama}

\address{Department of Physics, Faculty of Science,
Okayama University, Okayama 700-8530, Japan}

\begin{abstract}
Spectral properties for
the two-dimensional quantum $S=1$ $XY$ model
were investigated with the exact diagonalization method.
In the symmetry-broken phase,
there appear the massive Higgs 
and massless Goldstone 
excitations, which correspond to
the longitudinal 
and
transverse modes of the 
spontaneous magnetic moment, respectively.
The former excitation branch is embedded in the continuum of the latter,
and little attention has been paid to the details,
particularly, in proximity to the critical point.
The finite-size-scaling behavior is improved
by extending the interaction parameters.
An analysis of the critical amplitude ratio 
for these mass gaps is made.

\end{abstract}

\begin{keyword}

75.10.Jm 	
75.40.Mg 
05.50.+q , 
05.70.Jk 

\end{keyword}

\end{frontmatter}



\section{\label{section1} Introduction}

In the symmetry-broken phase,
the O$(2)$-symmetric system, such as the $XY$ model, exhibits
a massless Goldstone excitation, which corresponds to the transverse
modulation 
of the magnetic moment.
On the one hand,
the longitudinal mode, namely,
the Higgs excitation, is massive, embedded
in the continuum of the former; see Ref. \cite{Pekker15} for a review.
The O$(2)$- [equivalently, U$(1)$-] symmetric system is ubiquitous in nature,
and such a characteristic spectrum has been observed for 
a variety of
substances 
\cite{Ruegg08,Bissbort11,Endres12,Demsar99,Yusupov10,%
Lyons88,Parkinson69,Fleury70,Elliott69,Shastry90}.
The perturbation field (experimental probe)
should retain the O$(2)$ (axial) symmetry
\cite{Pekker15,Lindner10,Podolsky11,Huber07,Huber08};
otherwise, the contribution from the Goldstone excitations 
smears out the Higgs-mode branch \cite{Chubukov94,Sachdev99,Zwerger04,Dupuis09}.
(For instance, the chemical-potential modulation
for the bosonic system
does not conflict with the symmetry.)

Recent studies 
\cite{Podolsky12,Pollet12}
shed light on a universal character of the spectrum
in proximity to the phase transition,
especially, in $(2+1)$ dimensions;
it would be intriguing that the spectral
property is also under the reign of universality.
In $(3+1)$ dimensions, the criticality is described simply 
by the Ginzburg-Landau theory (Gaussian fixed point).
On the contrary, in $(2+1)$ dimensions,
the
spectral property is non-perturbative by nature.
In particular,
a universal amplitude ratio 
for the mass gaps
[see Eq. (\ref{amplitude_ratio}) mentioned afterward]
is arousing much attention recently.

In this paper,
we investigate the two-dimensional
quantum 
$S=1$-spin $XY$ model 
(\ref{Hamiltonian})
with the exact diagonalization method.
The method enables us to calculate
the low-lying level indexed by quantum numbers.
In order to suppress corrections to scaling,
we incorporate various types of interaction parameters 
in addition to
the ordinary nearest-neighbor 
ferromagnetic interaction $J_{NN}$.
Thereby, we investigate the universality for
the critical amplitude ratio
\begin{equation}
\label{amplitude_ratio}
m_H / \Delta  
\equiv
\frac{m_H(J_{NN})}{m_G(2J_{NN}^*-J_{NN})}
  ,  
\end{equation}
with the Higgs mass $m_H$, the Goldstone mass $m_G$, 
and the reflected gap $\Delta=m_G(2J_{NN}^*-J_{NN})$
with respect to the
critical point
$J^*_{NN}$; technical details and underlying physics are explained 
in the next section.
The amplitude ratio has been estimated as 
$m_H/\Delta=2.1(3)$ \cite{Gazit13a,Gazit13b} and 
$3.3(8)$ \cite{Chen13}
by means of the (quantum) Monte Carlo method.
According to the
recent elaborated
renormalization-group analyses,
the ratio was estimated as $2.4$ \cite{Rancon14},
$2.2$ \cite{Rose15},
and $1.67$  
\cite{Katan15}.

\begin{figure}
\includegraphics[width=100mm]{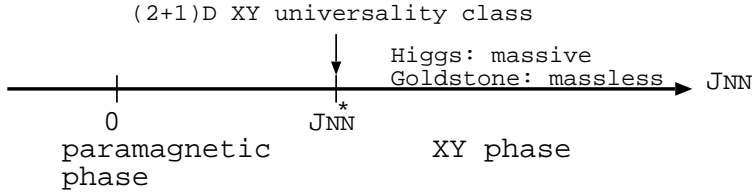}%
\caption{\label{figure1}
	A schematic phase diagram
 for the two-dimensional $S=1$ $XY$ model
(\ref{Hamiltonian})
is presented;
here, the parameter space is described by the formula (\ref{parameter_space}).
As the nearest-neighbor ferromagnetic interaction
$J_{NN}$ increases, a phase transition from the paramagnetic phase 
to the $XY$ phase occurs.
The critical behaviors for the Higgs- and Goldstone-mass gaps are the main concern.
}
\end{figure}

To be specific, 
we present 
the Hamiltonian for the $S=1$ $XY$ model
\cite{Nishiyama08}
\begin{eqnarray}
	{\cal H} & = &
-J_{NN} \sum_{\langle ij \rangle} (S^x_i S^x_j+S^y_i S^y_j)
-J_{NNN} \sum_{\langle\langle ij \rangle\rangle} (S^x_iS^x_j+S^y_iS^y_j)
                      \nonumber \\
\label{Hamiltonian}
& &
+D_{\Box} \sum_{[ijkl]} (S^z_i+S^z_j+S^z_k+S^z_l)^2
+D \sum_i (S^z_i)^2
   .
\end{eqnarray}
Here,
the quantum $S=1$-spin operator $ {\bf S}_i $
is
placed at each square-lattice point $i$.
The summations,
$\sum_{\langle ij\rangle}$, 
$\sum_{\langle\langle ij\rangle\rangle}$,
and 
$\sum_{[ijkl]}$,
run over all possible
nearest-neighbor, next-nearest-neighbor,
and plaquette spins,
respectively.
The parameters $(J_{NN},J_{NNN},D_\Box)$
are the corresponding coupling constants.
The parameter $D$ denotes the single-ion anisotropy.
We survey the coupling-constant subspace
\begin{equation}
\label{parameter_space}
(J_{NN},J_{NNN},D_\Box, D)=(J_{NN},J_{NNN}^* J_{NN}/J_{NN}^*
                 ,D_\Box^*,D^*)  ,
\end{equation}
parameterized by $J_{NN}$.
At $J_{NN}=J_{NN}^*$, the system undergoes
a phase transition;
a schematic phase diagram is presented in 
Fig. \ref{figure1}.
Here, the critical point
\begin{equation}
\label{FP_intro}
(J_{NN}^*,J_{NNN}^*,D_\Box^*,D^*)=
(
 0.158242810 160,
0.058561393564,
0.10035104389  ,
0.957)  ,
\end{equation}
was adjusted \cite{Nishiyama08}
to an IR fixed point with almost eliminated irrelevant interactions;
that is,
 the coupling constants $(J_{NN}^*,J_{NNN}^*,D_\Box^*)$
were determined through an approximative real-space renormalization group,
and the remaining one $D^*$ was finely tuned via the conventional 
finite-size-scaling analysis.
As shown in Eq. (\ref{Hamiltonian}),
the $S=1$-spin model allows us to incorporate various
interactions such as the single-ion anisotropy,
with which one is able to realize the $XY$-paramagnetic phase transition. 
In this sense, the extention of the magnetic moment to $S=1$ is
essential in our study.


The rest of this paper is
organized as follows.
In Sec. \ref{section2}, we present the simulation results.
Technical details are explained as well.
In Sec. \ref{section3}, we address the summary and discussions.

\section{\label{section2}Numerical results}

In this section, we present the simulation results.
To begin with, we explain the simulation technique.

\subsection{Simulation algorithm}

In this section,
we explain the simulation algorithm.
As mentioned in Introduction,
the $XY$ model
(\ref{Hamiltonian})
was simulated with the exact diagonalization method.
We implemented the
screw-boundary condition  \cite{Novotny90}
in order to 
treat a variety of system sizes $N=10,12,\dots,22$
($N$: number of constituent spins) systematically;
note that conventionally, the system size $N$ is restricted within $N=9,16,\dots$.
We adopt the
algorithm presented  in Sec. II of Ref. \cite{Nishiyama08}. 
The linear dimension $L$ is given by $L=\sqrt{N}$; note that the $N$ spins constitute a
rectangular cluster.

Thereby, we evaluated 
the mass gaps $m_H$ and $m_G$ via the following scheme.
The exact diagonalization method yields the low-lying energy levels 
$E_0<E_1<E_2 < \dots$ explicitly.
Each level $E_i(k,S^z_{tot})$
is specified by the momentum $k$ and 
the perpendicular magnetic moment $S^z_{tot}$;
in practice,
the numerical diagonalization was performed within the
subspace
$(k,S^z_{tot})$.
The Higgs- and Goldstone-mass gaps are characterized
by 
\begin{equation}
\label{Higgs_mass}
m_H=E_1 (0,0) -E_0 (0,0)
    ,
\end{equation}
and 
\begin{equation}
\label{Goldstone_mass}
m_G=E_0 (0,1) -E_0 (0,0)
   ,
\end{equation}
respectively.
The reflected gap $\Delta$ with respect to the critical point $J_{NN}^*$
is calculated by 
$\Delta=m_G(2J_{NN}^*-J_{NN})$.
The gap $m_G$ ($\Delta$) becomes massive in the paramagnetic ($XY$) phase $J_{NN}<J^*_{NN}$
($J_{NN} > J_{NN}^*$), and hence, the ratio $m_H/\Delta$ makes sense in the $XY$
phase.
The gap $\Delta >0$ is interpreted as the insulator gap
through regarding the ladder operators $S^{\pm}_i$ as the 
bosonic creation-annihilation operators.

\subsection{\label{section2_2}Scaling analyses of $m_{H,G}$}

In this section, we
investigate the scaling behaviors for
the mass gaps
$m_H$ (\ref{Higgs_mass}) and 
$m_G$ (\ref{Goldstone_mass}).

\begin{figure}
\includegraphics[width=100mm]{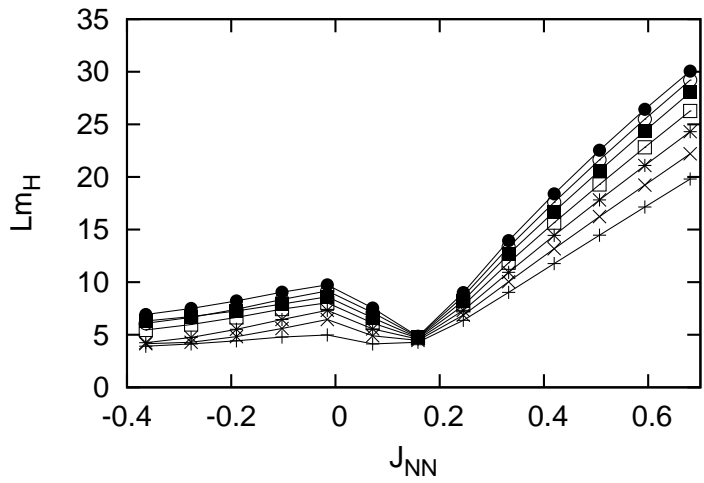}%
\caption{\label{figure2}
The
scaled Higgs mass $L m_H$
is plotted for various $J_{NN}$ and the system sizes of
($+$) $N(=L^2)=10$,
($\times$) $12$,
($*$) $14$,
($\Box$) $16$,
($\blacksquare$) $18$,
($\circ$) $20$,
and 
($\bullet$) $22$.
The scale-invariant point,
$J_{NN}^* \approx 0.15$, indicates the location of the critical point.
The Higgs gap opens in the $XY$ phase, $J_{NN}> J_{NN}^*$.
}
\end{figure}

In Fig. \ref{figure2},
we present the scaled Higgs gap $L m_H$ for the nearest-neighbor ferromagnetic
interaction
$J_{NN}$ and various system sizes $N=10,12,\dots,22$ ($L=\sqrt{N}$).
The data merge around $J_{NN} \approx 0.15$, indicating 
an onset of criticality;
note that the scaled energy gap should be scale-invariant 
at the critical point.
The location of the critical point is consistent with 
Eq. (\ref{FP_intro}).
The Higgs gap $m_H$ appears to open
in both the paramagnetic ($J_{NN} < J^*_{NN}$) and 
$XY$ ($J_{NN}>J^*_{NN}$ ) phases;
the latter case is our main concern, as mentioned in Introduction.

\begin{figure}
\includegraphics[width=100mm]{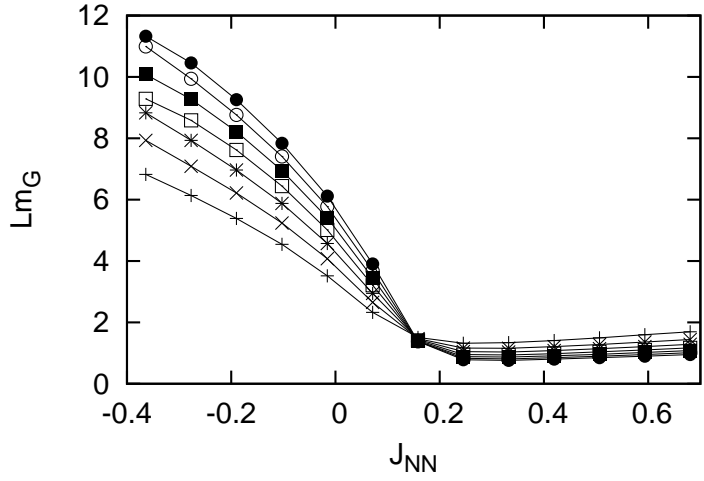}%
\caption{\label{figure3}
The
scaled Goldstone mass $L m_G$
is plotted for various $J_{NN}$ and the system sizes of
($+$) $N=10$,
($\times$) $12$,
($*$) $14$,
($\Box$) $16$,
($\blacksquare$) $18$,
($\circ$) $20$,
and 
($\bullet$) $22$.
The intersection point of the curves,
$J_{NN}^* \approx 0.15$, indicates the location of the critical point.
The Goldstone excitation is massless in the $XY$ phase,
$J_{NN}> J_{NN}^*$.
The mass $\Delta>0$ in the paramagnetic phase
is interpreted as the insulator gap in the boson language.
}
\end{figure}

In Fig. \ref{figure3},
we present the scaled Goldstone gap $L m_G$ for various $J_{NN}$
and $N=10,12,\dots , 22$.
The data indicate an onset of criticality around $J_{NN} \approx 0.15 $. 
The Goldstone gap closes in the $XY$ phase,
while it opens in the paramagnetic phase.
The latter gap is interpreted as the (bosonic)
insulator gap $\Delta$, setting a fundamental energy scale in this domain;
actually, the transition is interpreted as the superfluid-insulator
transition \cite{Pollet12}.
Hence, the ratio $m_H/\Delta$, Eq. (\ref{amplitude_ratio}), makes
sense in the $XY$ phase, and the criticality is investigated
in the next section.

\begin{figure}
\includegraphics[width=100mm]{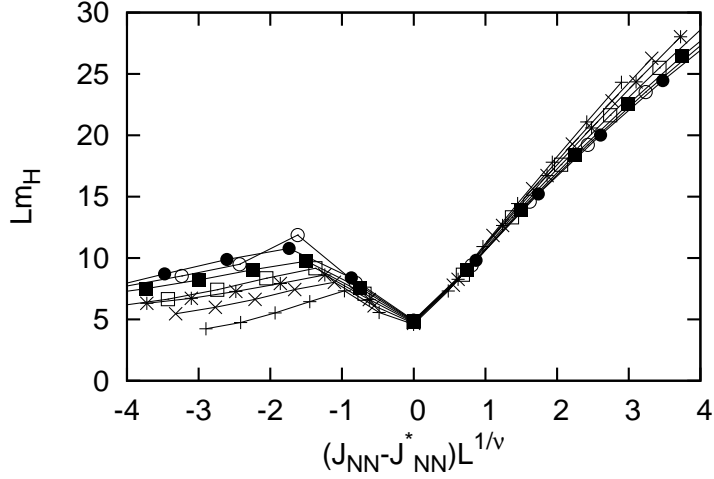}%
\caption{\label{figure4}
The scaling plot,
$(J_{NN}-J^*_{NN})L^{1/\nu}$-$L m_H$,
is presented for the system sizes
($+$) $N=10$,
($\times$) $12$,
($*$) $14$,
($\Box$) $16$,
($\blacksquare$) $18$,
($\circ$) $20$,
and 
($\bullet$) $22$.
The scaling parameters are set to 
$J_{NN}^*=0.158242810 160$ \cite{Nishiyama08}
and 
$\nu=0.6717$ \cite{Campostrini06}.
}
\end{figure}

\begin{figure}
\includegraphics[width=100mm]{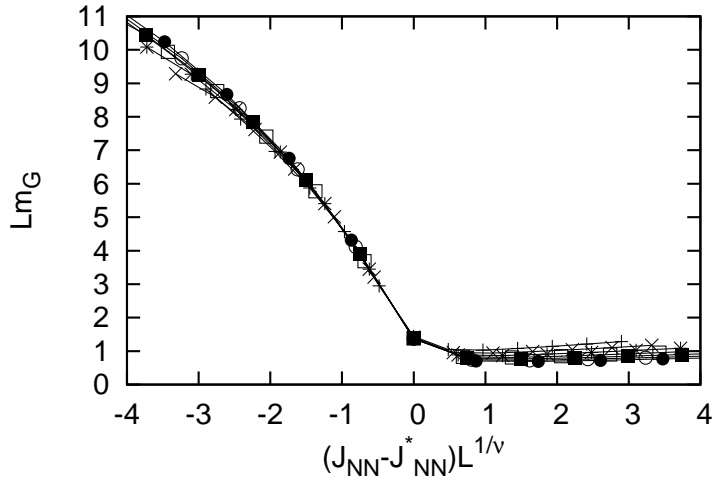}%
\caption{\label{figure5}
The scaling plot,
$(J_{NN}-J^*_{NN})L^{1/\nu}$-$L m_G$,
is presented for the system sizes
($+$) $N=10$,
($\times$) $12$,
($*$) $14$,
($\Box$) $16$,
($\blacksquare$) $18$,
($\circ$) $20$,
and 
($\bullet$) $22$.
The scaling parameters, 
$J_{NN}^*$ 
and 
$\nu$,
are the same as those of Fig. \ref{figure4}.
}
\end{figure}

In Fig. \ref{figure4},
 we present the scaling plot,
$(J_{NN}-J_{NN}^*)L^{1/\nu}$-$L m_H $, for various system sizes $N=10,12,\dots,22$.
Here, the scaling parameters are set to 
$J_{NN}^*=0.158242810 160 $ [Eq. (\ref{FP_intro})] and 
$\nu=0.6717$ \cite{Campostrini06,Burovski06}.
The data appear to collapse into a scaling curve satisfactorily.
Similarly, in Fig. \ref{figure5},
we present the scaling plot,
$(J_{NN}-J^*_{NN})L^{1/\nu}$-$L m_G $, for various system sizes $N=10,12,\dots,22$;
the scaling parameters are the same as those of Fig. \ref{figure4}.

We address a few remarks.
First, the data in Figs. \ref{figure4} and \ref{figure5}
collapse into the scaling curves satisfactorily.
Such a feature indicates that 
corrections to scaling are almost negligible owing to
the fine adjustment 
\cite{Nishiyama08}
of the coupling constants to Eq. (\ref{FP_intro}).
Because the tractable system size with the exact diagonalization method
is severely restricted, it is significant to accelerate the convergence
to the scaling limit.
Second,
the scaling parameters, $J_{NN}^*$ 
and $\nu$, are 
taken from the literatures,
Refs.
\cite{Nishiyama08} and \cite{Campostrini06}, respectively.
That is, there are no adjustable {\it ad hoc} parameters in the present 
scaling analyses.
Last, as demonstrated in Figs. \ref{figure2} and \ref{figure3},
both mass gaps $m_{H,G}$ possess an identical scaling dimension.
Hence, the amplitude ratio (\ref{amplitude_ratio})
makes sense, and the criticality is explored in the next section.

\subsection{Analysis of the amplitude ratio $m_H/\Delta$}

In this section, encouraged by the findings in 
Sec. \ref{section2_2},
we turn to the analysis of the
amplitude ratio $m_H/\Delta$, Eq. (\ref{amplitude_ratio}).

\begin{figure}
\includegraphics[width=100mm]{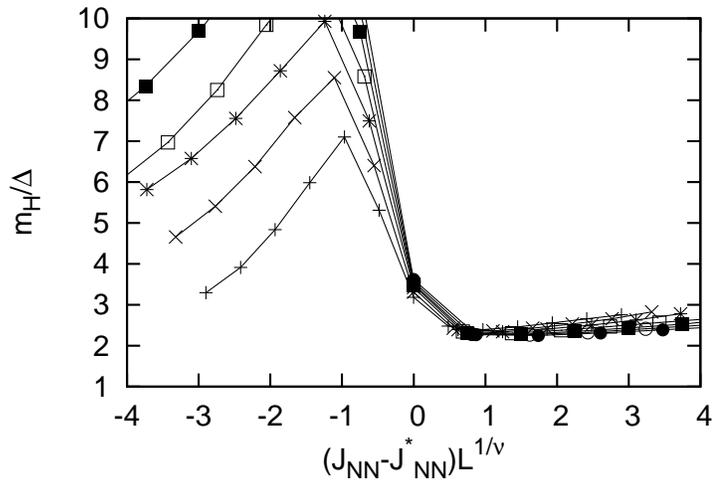}%
\caption{\label{figure6}
The scaling plot,
$(J_{NN}-J^*_{NN})L^{1/\nu}$-$m_H/\Delta$,
is presented for the system sizes
($+$) $N=10$,
($\times$) $12$,
($*$) $14$,
($\Box$) $16$,
($\blacksquare$) $18$,
($\circ$) $20$,
and 
($\bullet$) $22$.
The scaling parameters, 
$J_{NN}^*$ 
and 
$\nu$,
are the same as those of Fig. \ref{figure4}.
A plateau appears in the $XY$ phase,
indicating a universal character of the amplitude ratio.
}
\end{figure}

In Fig. \ref{figure6},
we present the scaling plot,
$(J_{NN}-J_{NN}^*)L^{1/\nu}$-$m_H/ \Delta$,
for $N=10,12,\dots,22$; 
here, the scaling parameters, $J_{NN}^*$ and $\nu$,
are the same as those of Fig. \ref{figure4}.
In the $XY$ phase, $J_{NN}-J_{NN}^* > 0$, the amplitude ratio exhibits a plateau
for an appreciable range of $J_{NN}$.
Such a feature clearly indicates that 
the amplitude ratio is a universal constant in this domain.

\begin{figure}
\includegraphics[width=100mm]{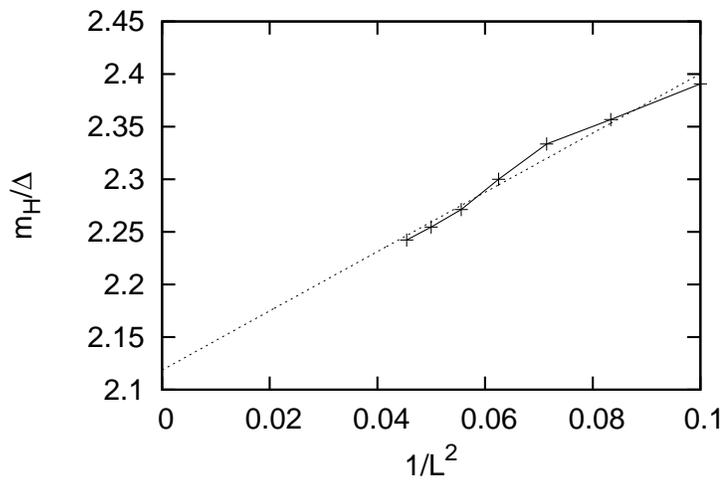}%
\caption{\label{figure7}
The mass-gap amplitude ratio 
$m_H/\Delta$ (\ref{amplitude_ratio_calculation})
is plotted for $1/L^2$ ($N=10,12,\dots,22$).
The least-squares fit to the data yields 
$m_H/\Delta=2.119(13)$ in the thermodynamic limit
$L \to \infty$.
A possible extrapolation error is considered in the text.
An oscillatory deviation (slight bump around $L^2 \approx 16$) is 
an artifact of the
screw-boundary condition \cite{Novotny90}.
}
\end{figure}

Upon close inspection, the plateaux in Fig. \ref{figure6}  are curved concavely.
The shallow bottom locates at 
$J_{NN}=\bar{J}_{NN}$, satisfying 
$\partial_{J_{NN}} (m_H/\Delta)|_{J_{NN}=\bar{J}_{NN}}=0$
for each system size $N$.
We regard the bottom height 
\begin{equation}
\label{amplitude_ratio_calculation}
m_H/\Delta |_{J_{NN}=\bar{J}_{NN}} ,
\end{equation}
as an indicator for $m_H/\Delta$.
The amplitude ratio (\ref{amplitude_ratio_calculation}) 
is plotted for $1/L^2$ 
[$N(=L^2)=10,12,\dots,22$]
in Fig. \ref{figure7}.
The least-squares fit to the data yields an estimate 
$m_H/\Delta=2.119(13)$
in the thermodynamic
limit
$L \to \infty$.
As a reference, a similar analysis was performed with the abscissa scale
replaced with $1/L$, and we arrived at $m_H/\Delta = 1.923(17)$.
The discrepancy $\approx 0.2$ between these estimates
appears to dominate the least-squares-fit error $\approx 0.02$,
and the discrepancy may 
indicate an ambiguity as to the extrapolation
(systematic error).
Regarding it as a possible systematic error,
we estimate the amplitude ratio as
\begin{equation}
m_H/\Delta = 2.1(2)
  .
\end{equation}

A comment may be in order,
the series of data in Fig. \ref{figure7}
appear to be oscillatory;
actually, we observe a slight bump around $N(=L^2)\approx 16$.
Such an oscillatory behavior is an artifact of  the screw-boundary condition 
\cite{Novotny90}, rendering an ambiguity as to the extrapolation to
$L\to\infty$.
The ambiguity appears to be bounded by the above-mentioned error margin,
which is
estimated by performing two independent extrapolation schemes.

\section{\label{section3}
Summary and discussions}

The critical behavior 
of $m_{H,G}$
was investigated
for the two-dimensional quantum $S=1$ $XY$ model 
(\ref{Hamiltonian})
by means of the numerical diagonalization method
\cite{Novotny90,Nishiyama08}.
The interaction parameters were adjusted to
Eq. (\ref{parameter_space})
in order to suppress corrections to scaling \cite{Nishiyama08}.
As a consequence,
the data
(Figs. \ref{figure4} and \ref{figure5})
collapse into the scaling curves
satisfactorily, indicating that 
the data already enter the scaling regime.
Thereby, 
we confirm a universal character for 
the mass-gap ratio (Fig. \ref{figure6}),
and estimate 
the amplitude ratio 
as $m_H/\Delta =2.1(2)$.

As mentioned in Introduction,
the amplitude ratio
has been
estimated with
the (quantum) Monte Carlo method,
$m_H/\Delta=2.1(3)$ \cite{Gazit13a,Gazit13b} and
$3.3(8)$ \cite{Chen13},
as well as 
the renormalization-group approaches,
$2.4$ \cite{Rancon14},
$2.2$ \cite{Rose15},
and $1.67$ 
\cite{Katan15}.
According to the
Ginzburg-Landau (mean-field) theory,
the amplitude ratio should be
$m_H/\Delta=\sqrt{2}$.
Clearly, the spectral property reveals a
notable deviation from that anticipated from the mean-field theory;
the Ising counterpart was 
studied in Ref. \cite{Dusuel10}.
In this respect,
detailed analyses of
other
spectral properties such as the AC conductivity
\cite{Gazit13b,Chen14}
would be desirable. A progress
toward this direction is left for the future study.

\section*{Acknowledgment}
This work was supported by a Grant-in-Aid
for Scientific Research (C)
from Japan Society for the Promotion of Science
(Grant No. 25400402).

\end{document}